\documentclass[aps,prb,twocolumn,superscriptaddress,showpacs,floatfix,amsmath,amssymb]{revtex4-1}
\usepackage{graphicx}
\usepackage{dcolumn}
\usepackage{bm}
\usepackage{times}
\usepackage{color}    
\usepackage{hyperref}  
\usepackage{mciteplus} 
\usepackage{float} 
\def\etal{{\em{et al}}}

\begin{document}

\title{Electronic structure and X-ray spectroscopy of Cu$_{2}$MnAl$_{1-x}$Ga$_{x}$} 

\author{D. P. Rai}
\email[E-mail: ]{dibyaprakashrai@gmail.com}
\affiliation{Department of Physics, Pachhunga University College, Aizawl, 796001, India}
\author{C. E. Ekuma}
\email[E-mail: ]{cekuma1@gmail.com}
\affiliation{Computational Material Physics Group, Department of Physics, University of Port Harcourt, PMB 51001, Rivers, Nigeria}
\author{A. Boochani}
\affiliation{Department of Physics, Kermanshah Branch, Islamic Azad University, Kermanshah, Iran}
\author{S. Solaymani}
\affiliation{Department of Physics, Science and Research Branch, Islamic Azad University, Tehran, Iran.}
\author{R. K. Thapa}
\affiliation{Department of Physics, Mizoram University, Aizawl, 796009, India}

\begin{abstract}
We explore the electronic and related properties of Cu$_{2}$MnAl$_{1-x}$Ga$_{x}$ with a first-principles, relativistic multiscattering Green function approach. We discuss our results in relation to existing experimental data and show that the electron-core hole interaction is essential for the description of the optical spectra especially in describing the X-ray absorption and magnetic circular dichroism spectra at the L$_{2,3}$ edges of Cu and Mn.
\end{abstract}

    
\maketitle

\section{\label{sec:level1}Introduction }
The Heusler compounds are increasingly being explored for diverse modern device applications due to their fascinating properties not limited to martensitic phase transformations, half-metallicity, inverse magnetocaloric effect, magnetic shape memory, thermoelectric capabilities, and high Curie temperature\cite{degroot,Ishida,Burnstein2001,10.1063/1.2149185,*10.1063/1.3609075,Fert08,*ANIE200801093,Kudr02,Gala03,Rai2017,Schw86,Kim03,Liu04,Yama07,dpr2016,dpr2010,Koba98,Galanakis,Antonov}. The origin of the multi-properties has been ascribed to chemical ordering, crystal field splitting, electron correlation, exchange-coupling, and the $d$-bandwidth~\cite{Burnstein2001}. Doping these compounds catalyzes an atomistic coupling that gives rise to high charge carriers, giant magneto-optical (Zeeman) effect, and magnetotransport properties, essential for multifunctional device applications~\cite{Chiba2003,Chiba2008}. For e.g., high Curie temperatures $\sim$840 K and 1100 K has been reported in Fe$_3$S/GaAs hybrid structure~\cite{Herf03} and Co$_2$FeSi~\cite{Wurm06}, respectively. Experimental insight into the atomistic scale of these materials can be achieved using magneto-optical spectroscopy, e.g, the X-ray absorption spectroscopy (XAS) and X-ray magnetic circular dichroism  (XMCD)~\cite{Ando,Thole,Miyokawa,Jungwirth}, which provide direct fingerprint of the local spin and orbital magnetic moments of transition metals in the case of the L$_{2,3}$ absorption edges. Computational techniques employing the varying degree of theoretical complexity to compute the XAS and XMCD~\cite{PhysRevLett.70.694,Laan,Jo,Piamonteze,hedman2007x} has recently become possible.

The properties of materials can be tuned across the electromagnetic spectra via defect-engineering. Doping plays a crucial role in modulating the charge carriers, which in turn can control, e.g., the Curie temperature~\cite{Chiba2003} and the magnetic anisotropy~\cite{Chiba2008} in field effect devices. Unlike other transition metal-based Heusler compounds, Cu$_{2}$MnAl is unique as it deviates from the Slater-Pauling rule~\cite{Erb2010,Geiersbach}. Their complicated magnetic structure could be attributed to Friedel oscillations produced by a potential scattering, which induces indirect exchange interaction among the magnetic atoms that strongly depends on the conduction electron concentration~\cite{PhysRevB.77.064417} (herein Ga). Also, due to the small distance between the Mn and Cu atoms, there is considerable overlap of the 3$d$ wavefunctions (direct exchange). An intricate coupling of two or more of these effects can induce ferromagnetism, Pauli paramagnetism, helimagnetism, antiferromagnetism, or heavy-fermion behavior in full-Heusler alloys~\cite{Burnstein2001,Guo1998,0953-8984-10-5-011,*0022-3727-37-15-001,felser2015heusler}. In the case of Cu$_{2}$MnAl, it orders ferromagnetically with $L2_1$-type structure, a Curie temperature T$_C$ $\sim$603 K, and a rather large magnetic moment $\sim$3.30 $\mu_B$~\cite{Rai-thapa,Oxley}.  Further, Cu$_{2}$MnAl is considered to be an ultra flexible system for studying the effect of doping on the magnetic properties. An  $sp$-electron concentration can be used to explore the variation of localized moment at 3$d$ orbital and magnetic properties. 

The effect of disorder in Cu$_{2}$MnAl due to partial replacement by impurity atoms at various sites have been studied~\cite{Konoplyuk,Bang,Zhang}. Singh \etal~\cite{Singh} in their experimental work, reported that an increase in Ga concentration at Al site gives rise to $\gamma$-Cu$_{9}$Al$_{4}$ and Cu$_{2}$MnAl phases. The authors also reported that an annealed Cu$_{50}$Mn$_{25}$Al$_{25-x}$Ga$_{x}$ ($x=0$,8) at 903 K for 30 h leads to a $\beta$-Mn and $\gamma$-Cu$_{9}$Al$_{4}$ phases. However, we are not aware of any complementary first-principles computational study of the effect of Ga doping at Al site.  In particular, herein, we investigate the fundamental relation of the physical ordering on the electronic and the magnetic properties of  Cu$_{2}$MnAl$_{1-x}$Ga$_{x}$ using a first-principles,  relativistic multiple scattering Green function approach by computing the XAS and XMCD at the L$_{2,3}$ edges of Cu and Mn, respectively.

\section{Computational details}
Cu$_{2}$MnAl is a prototype full-Heusler alloy crystallizing in the $L2_1$ structure~\cite{Heusler,*Buschow} with space group $Fm3m$. The unit cell consists of four interpenetrating face-centered-cubic sublattices with the Wyckoff positions as indicated between parentheses: Cu1: (1/4,1/4,1/4), Cu2: (3/4,3/4,3/4), Mn: (1/2,1/2,1/2), and Al: (0,0,0).  We used a well-converged basis set with dense Brillouin zone sampling, which are needed for the absorption properties. For this purpose, we used a uniform $30\times30\times30$ Monkhorst-pack grid. We used the full-potential linearized augmented planewave as implemented in the spin-polarized, relativistic Korringa-Kohn-Rostoker (SPR-KKR)~\cite{Ebert1996,*Ebert-kkr} within the density functional theory (DFT)\cite{Hohenberg1964,*Kohn1965} using the generalized gradient approximation of Perdew-Burke-Ernzerhof~\cite{PhysRevLett.77.3865} to treat the electron exchange-correlation. The relativistic effect is taken into account via scalar relativistic approximation and the disorder incorporated by randomly substituting Ga atoms at Al sites using the site occupancy of the dopant atom. We neglect the lattice distortion due to addition of impurity and use the lattice constant $a$=5.957 (\AA), which is obtained from our previous work~\cite{Rai-thapa}. We carefully checked the convergence of our results on various computational parameters, such as energy cutoffs and Brillouin zone sampling. We estimated the strength of the magnetic interaction by calculating the exchange coupling constants $J_{ij}$ between two atoms at $i$ and $j$ sites as a function of distance given by 
\begin{equation}
J_{ij} = \frac{1}{4\pi} \int^{E_F} dE \,  \Im \, Tr_L\{ \Delta_i \hat{T}^{ij}_\uparrow   \Delta_j \hat{T}^{ji}_\downarrow \} 
\label{eq:eq1}
\end{equation}
as implemented within the  classical Heisenberg model~ \cite{Liechtenstein}, where ${\Delta}_{i}=t_{i\uparrow}^{-1}-t_{j\downarrow}^{-1}$,  $t_{\uparrow\downarrow}^{-1}$ is the atomic $t$-matrix of the magnetic impurities at site $i$ for the spin up/down states, $\hat{T}_{\uparrow\downarrow}^{ij}$ is the scattering path operator between $ij$ sites  for the spin up/down states, and $Tr_{L}$ is the trace over the orbital variables. Then, by using mean field approximation, a Curie temperature  (T$_{C}$) can be estimated as 
\begin{equation}
k_{B}T_{C} \langle s_i \rangle =\frac{2}{3} x \sum_{j,r\neq0}J_{ij}^{0r} \langle s_j \rangle, 
\end{equation}
where $x$ is the concentration of impurity, $\langle s_j \rangle$ is the average $j$ component of the unit vector $s_r^j$ in the direction of the magnetic moment at site ($j,r$), and $k_{B}$ is the Boltzmann constant. The calculated value of T$_{C}$ from MFA is always overestimated as compared to experimental one, due to the insufficient description of the magnetic percolation effect \cite{Sato,PhysRevLett.93.137202}. 

\begin{figure*}[t!]
    \centering
      \includegraphics[width=5.40cm]{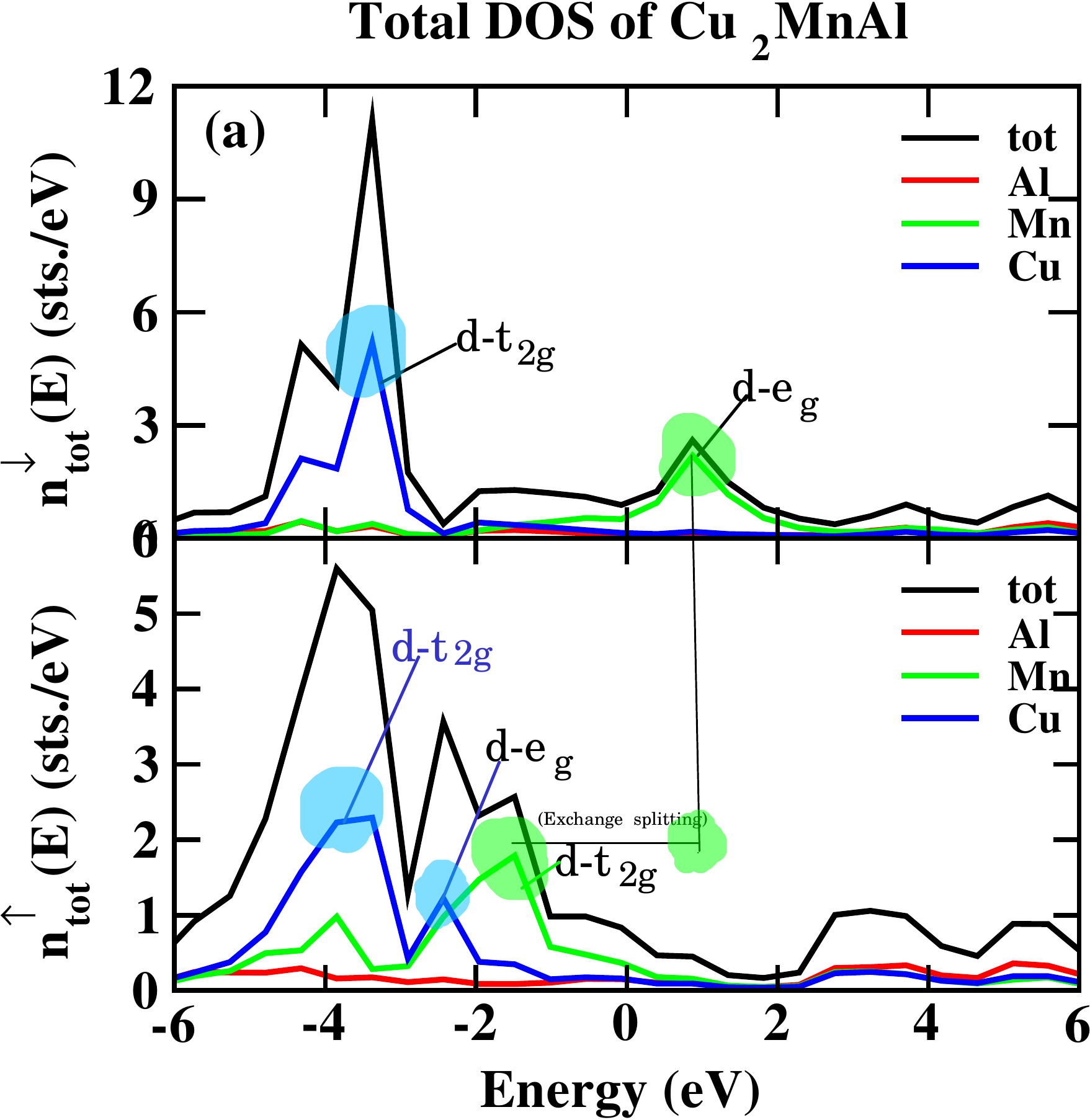}
    \includegraphics[width=5.40cm]{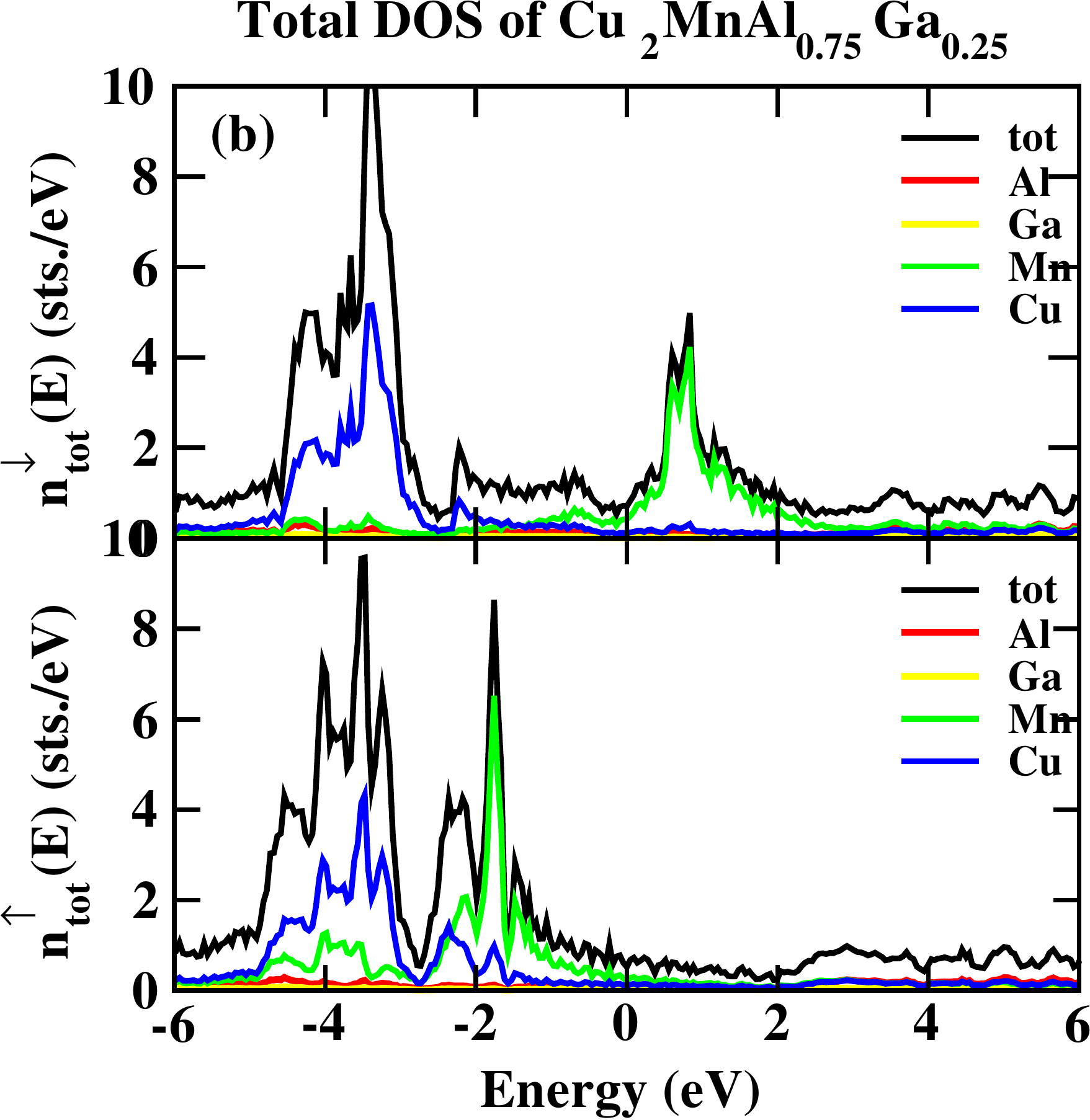}
    \includegraphics[width=5.40cm]{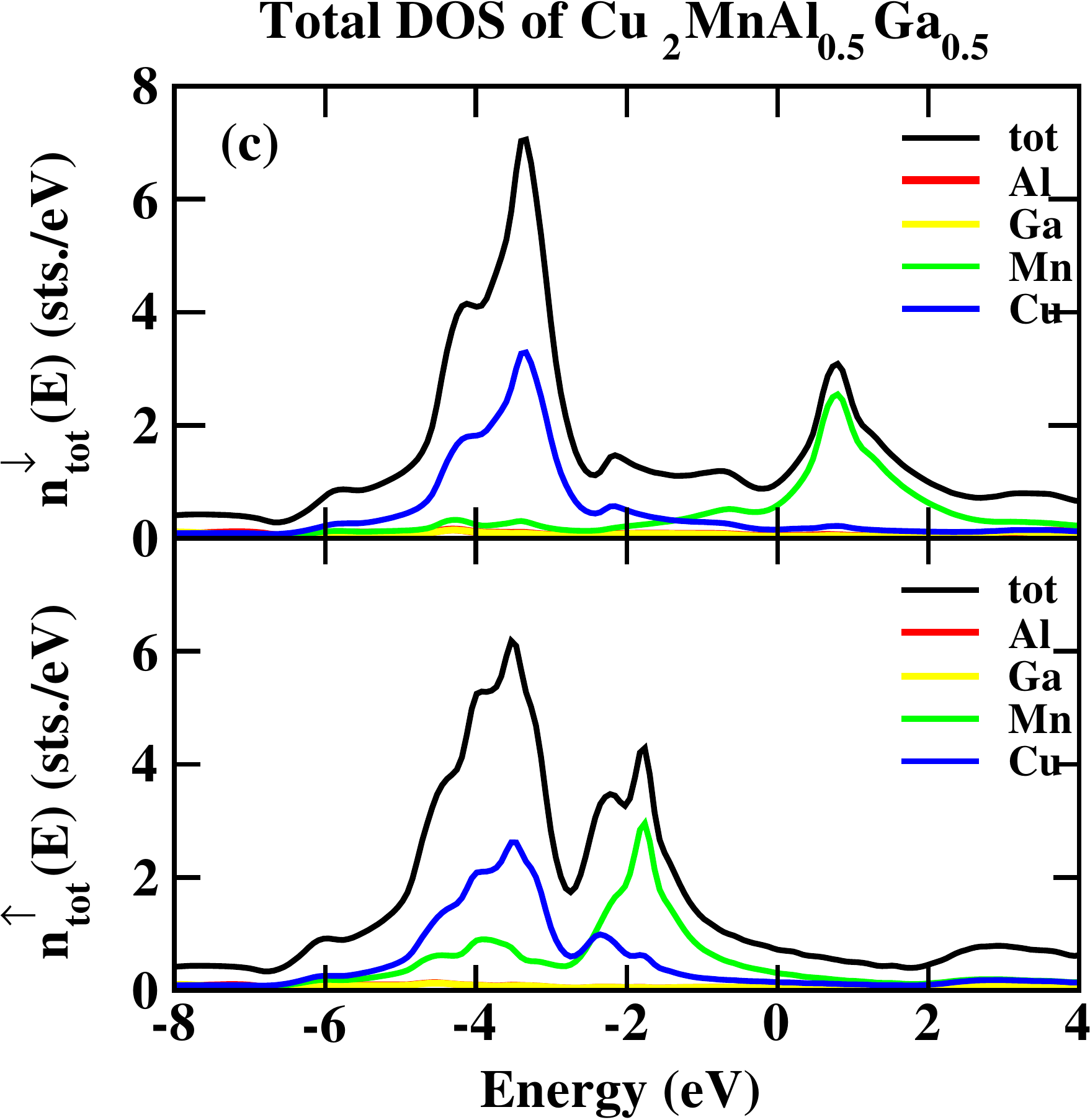}
      \includegraphics[width=5.40cm]{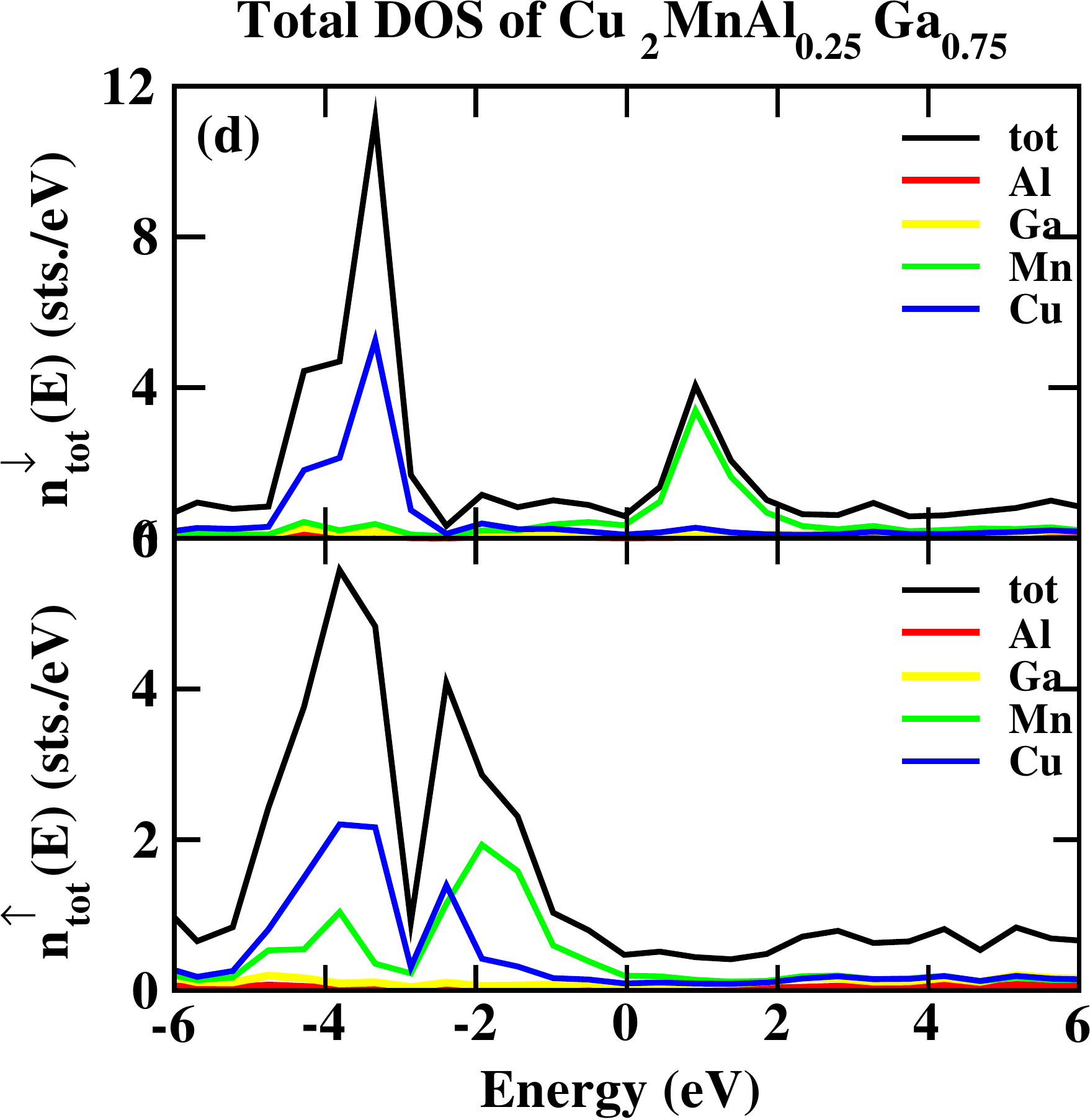}
        \includegraphics[width=5.40cm]{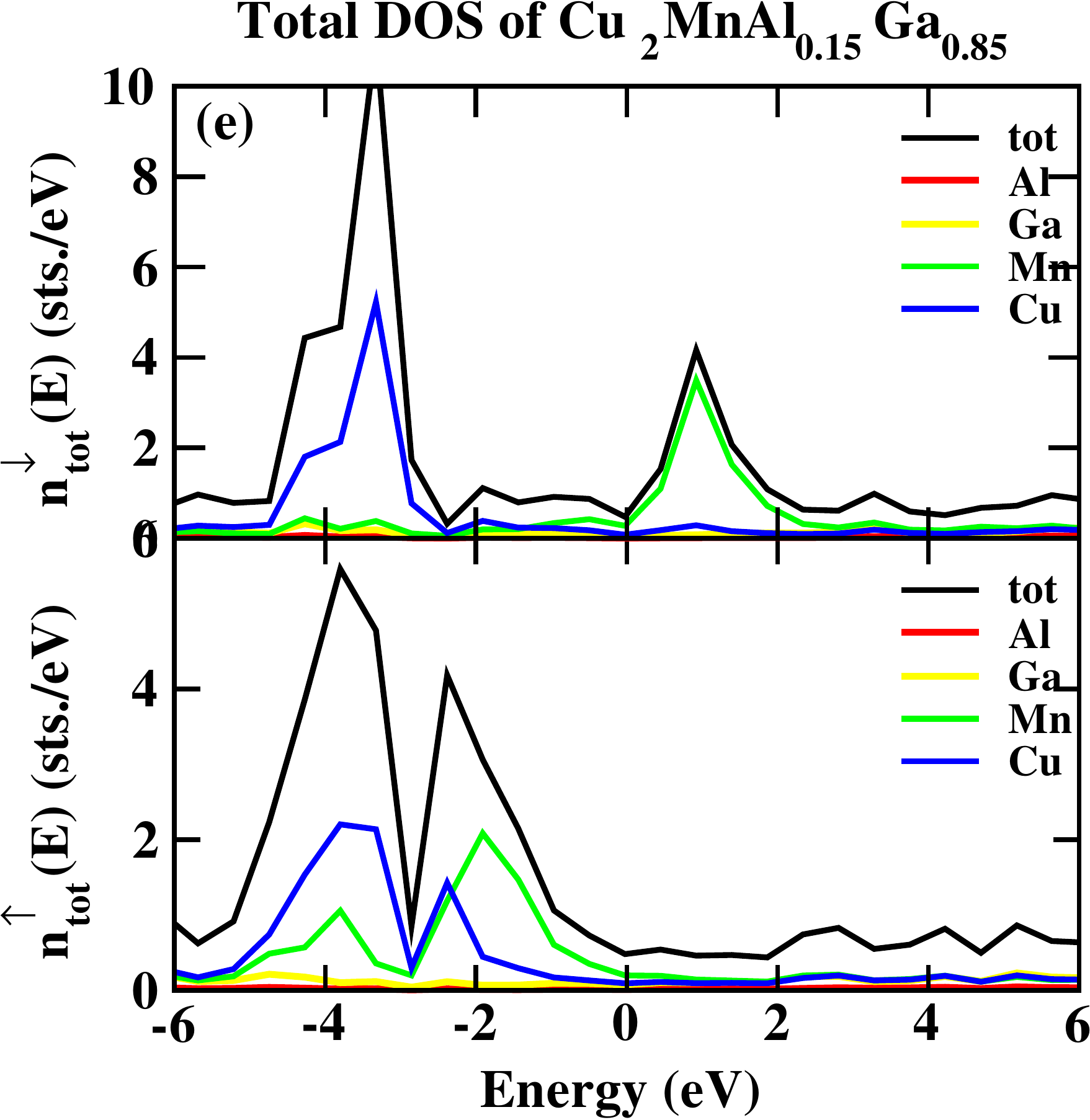}
          \includegraphics[width=5.40cm]{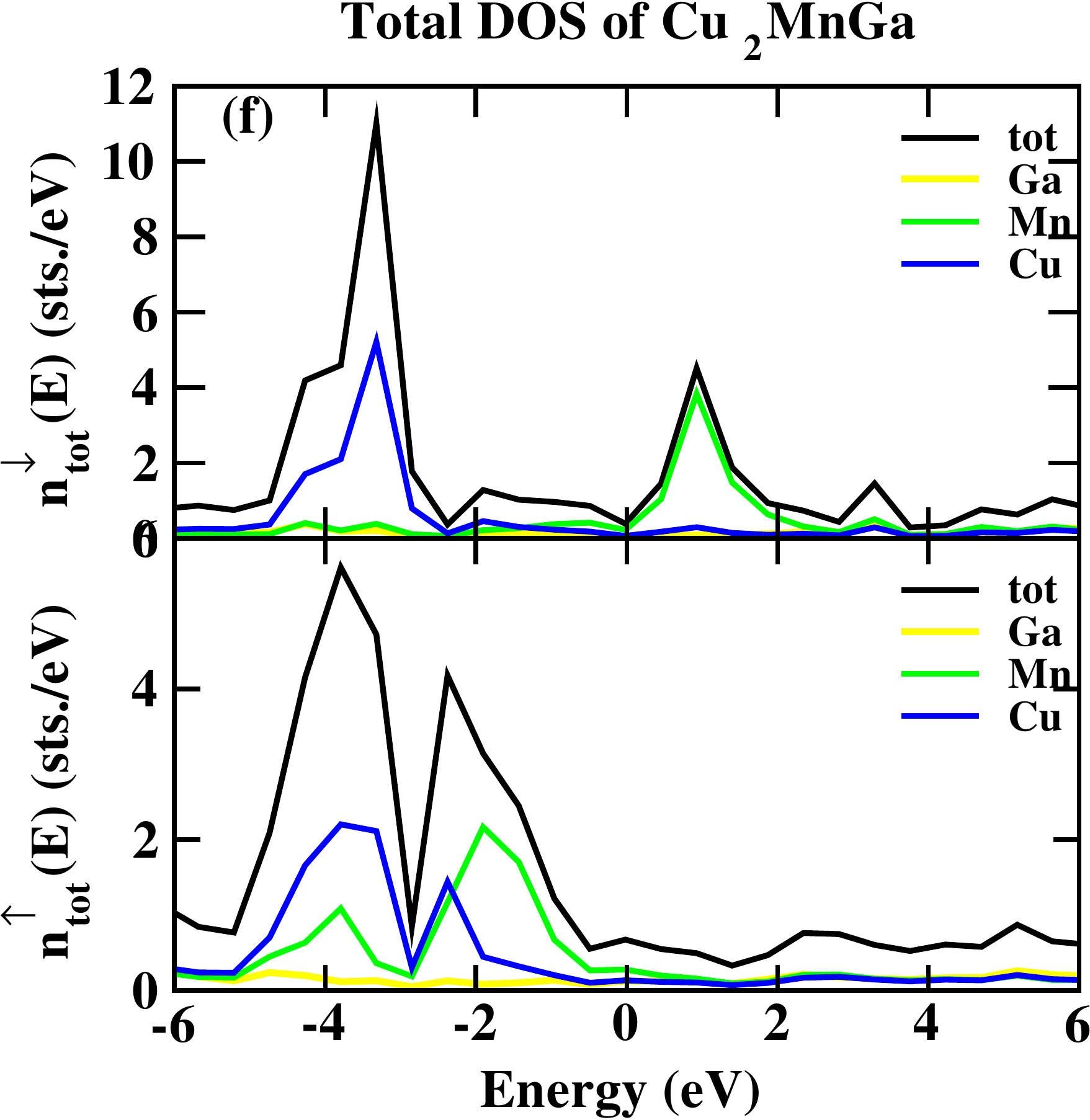}
    \caption{The total density of states with the corresponding projected density of states for Cu, Mn, Al, and Ga for Cu$_{2}$MnAl$_{1-x}$Ga$_{x}$ at various Ga defect concentrations $x$: (a) x=0.0 (b) x=0.25 (c) x=5.0 (d) x=0.75 (e) x=0.85 and (f) x=1.0 obtained with the first-principles, relativistic multiscattering KKR approach for both spin up ($\uparrow$) and spin down ($\downarrow$) channel.}    
       \label{fig:CMA-dos}
\end{figure*}

\section{Result and Discussion}

\subsection{Electronic and magnetic properties}
We show in Fig.~\ref{fig:CMA-dos} the total density of states of Cu$_{2}$MnAl$_{1-x}$Ga$_{x}$ at various Ga defect concentration and spin configurations. For the pristine case ($x=0$), the majority of the valence bands are from Mn$-d$ whereas that of the conduction bands emanate predominantly from Cu$-d$ states. Our calculations show that the Al-$sp$ states are fully localized and mostly contributed in the valence region. We also find a degenerated $d$-orbital: $d-e_{g}$ ($d_{z^2}$$+$$d_{x^2-y^2}$) and $d-t_{2g}$ ($d_{xy}$$+$$d_{yz}$$+$$d_{zx}$) in agreement with previous work~\cite{Rai-thapa}. In a typical X$_{2}$YZ-type full-Heusler alloy compound, the formation of energy band gap has been attributed to a hybridization between the $d-d$ orbitals of the transition metals~\cite{Galanakis}. However, in Cu$_{2}$MnAl, the Cu$-d$ orbital is completely occupied (Cu$-d^{10}$) and localized below the Fermi energy E$_{F}$. Therefore, Cu$-d$ does not take part in the $d-d$ hybridization with Mn$-d$ orbital to give rise to bonding and anti-bonding states. As a result, there is no band gap at E$_{F}$ in both the spin channels.

As we vary the concentration of Ga doping concentration [c.f. Fig.~\ref{fig:CMA-dos}(a-f)], we observe that the Cu-$d$ states are fully occupied and shifted in the valence region with a negligible magnetic moment $\sim$ 0.046 $\mu_B$. This rather small magnetic moment depicts that Cu is non-ferromagnetic in Cu$_{2}$MnAl$_{1-x}$Ga$_{x}$. The two prominent peaks at -2.5 and -3.5 eV are due to the Cu-$d_{e_{g}}$ and Cu-$d_{t_{2g}}$ states, respectively. The occurrence of the similar peak due to Mn-$d_{e_{g}}$ in spin up channel may give rise to the $d-d$ bonding of Cu and Mn atom. The Mn$-d$ states show anti-symmetric density of states with two prominent peaks at 1.5 eV (spin down) and -2.5 eV (spin up) due to the $d-e_{g}$ and $d-t_{2g}$ states, respectively. The sharp and prominent peaks due to Mn-$d$ and Cu-$d$ states are highlighted with green and blue shades, respectively. The antisymmetric density of states between Mn-$d_{e_{g}}$ and Mn-$d_{t_{2g}}$ is due to exchange splitting $\triangle_{xc}$, which signifies magnetic moment at Mn sites $\sim$3.38 $\mu_B$ from our calculations. Our calculated spin/orbital magnetic moment values (c.f.  Table \ref{sec:level1} for details) are in accord with previously reported values of 0.033/0.009 and 3.36/0.20 $\mu_B$ for Cu and Mn, respectively~\cite{Krume}. Analyzing the projected density of states further for the various Ga doping in the Al site [c.f. Fig.\ref{fig:CMA-dos}(a-f)], we observe overall nontrivial shifting of the $d$-states towards lower energy leading to the widening of the Mn-$d_{e_{g}}$ and Mn-$d_{t_{2g}}$ in energy scale. This trend enhances the exchange splitting and hence the magnetic moment (c.f. Table~\ref{sec:level1}). This trend seems to support the idea that in disordered A2-structure, Mn bonding-antibonding interactions undergo strong antiferromagnetic direction exchange interaction in Cu$_{2}$MnAl~\cite{Geiersbach}.

\begin{figure}
    \centering
    \includegraphics[trim = 0mm 0mm 0mm 0mm,scale=0.20,keepaspectratio,clip=true]{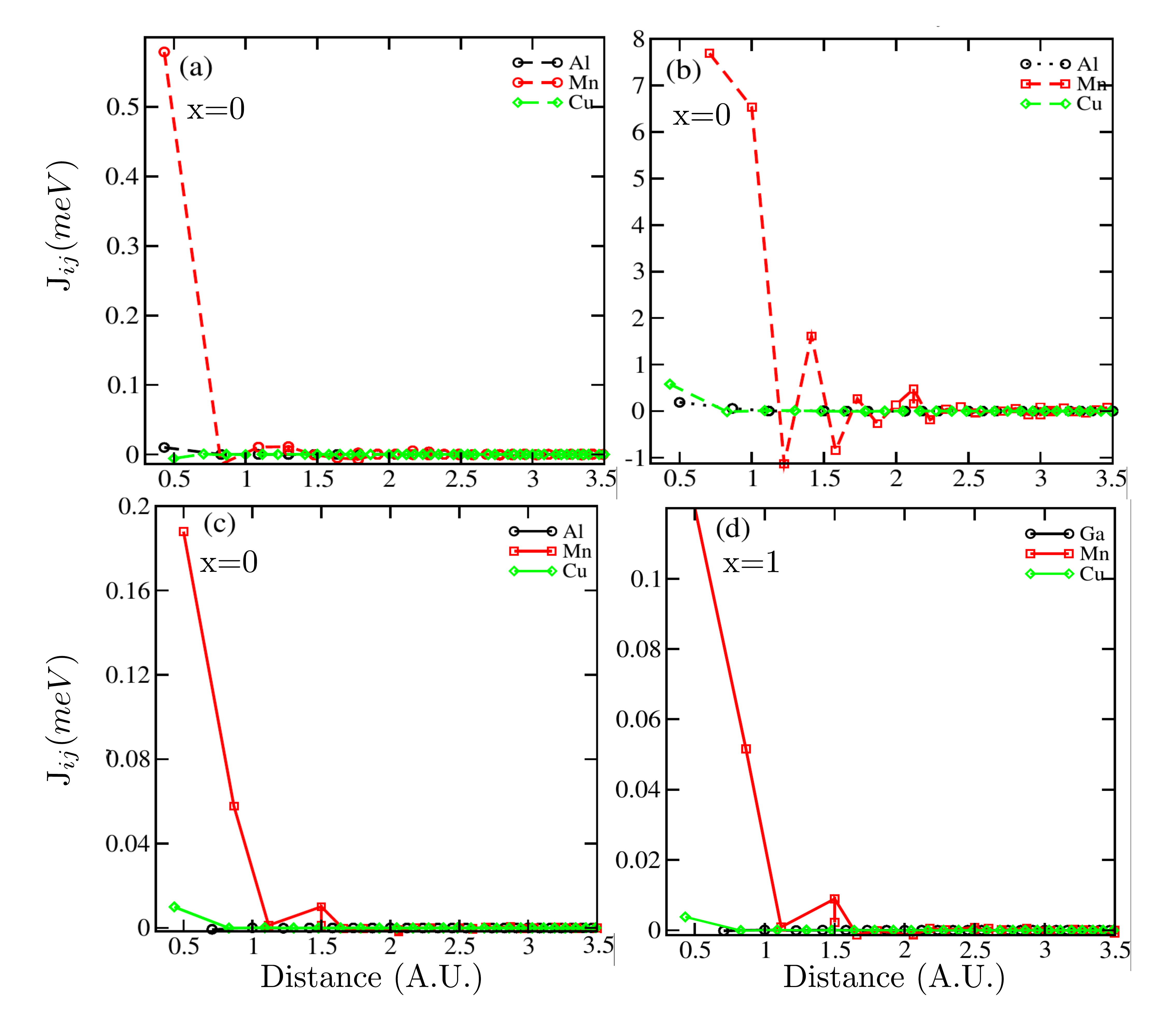}
    \caption{The exchange Coupling (J$_{ij}$) in Cu$_2$MnAl$_{1-x}$Ga$_x$ for (a) Cu-, (b) Mn-, (c) Al-, and (d) Ga-center as obtained using the first-principles, relativistic multiscattering KKR approach.}
    \label{fig:exc}
\end{figure}

To gain deeper insight into the role of Ga doping on the magnetic properties of Cu$_2$MnAl$_{1-x}$Ga$_x$ ($x=$0.0 and 1.0), we show in Fig.~\ref{fig:exc}(a-d) the exchange coupling constant $J_{ij}$ as a function of the distance between two atoms obtained using Eq.~\ref{eq:eq1}. At small interatomic distances, we observe strong $J_{ij}$, which decays exponentially as the distance between the atoms increases (c.f. Fig~\ref{fig:exc}). We attribute this to the rather exponentially decay of the wavefunction associated with deeper states in the spectra. The high value of short-ranged interactions is a signature of promising functional properties such as Curie temperature, electric conductivity, and magnon dispersion. As evident from Fig.~\ref{fig:exc} (a-d), the neighbors interacting with Mn atom lead to higher $J_{ij}$. In all cases, the Mn-Mn interaction is the strongest with a maximum value of $J_{ij}$ in short-range and fluctuates at an average range. We attribute this high value of $J_{ij}$ to the active participation of delocalized Mn-$d$ electrons. Observe that the $J_{ij}$ is smallest for $x=1$. This is in agreement with earlier results, which show that $J_{ij}$ decreases in disordered Fe(V,Mn)Sb, with an increase in impurity concentration, a fingerprint of a ferromagnetic system due to weak superexchange interaction~\cite{Fukushima}.  

\begin{figure}[htb]
    \centering
    \includegraphics[width=4.5cm,height=3.5cm]{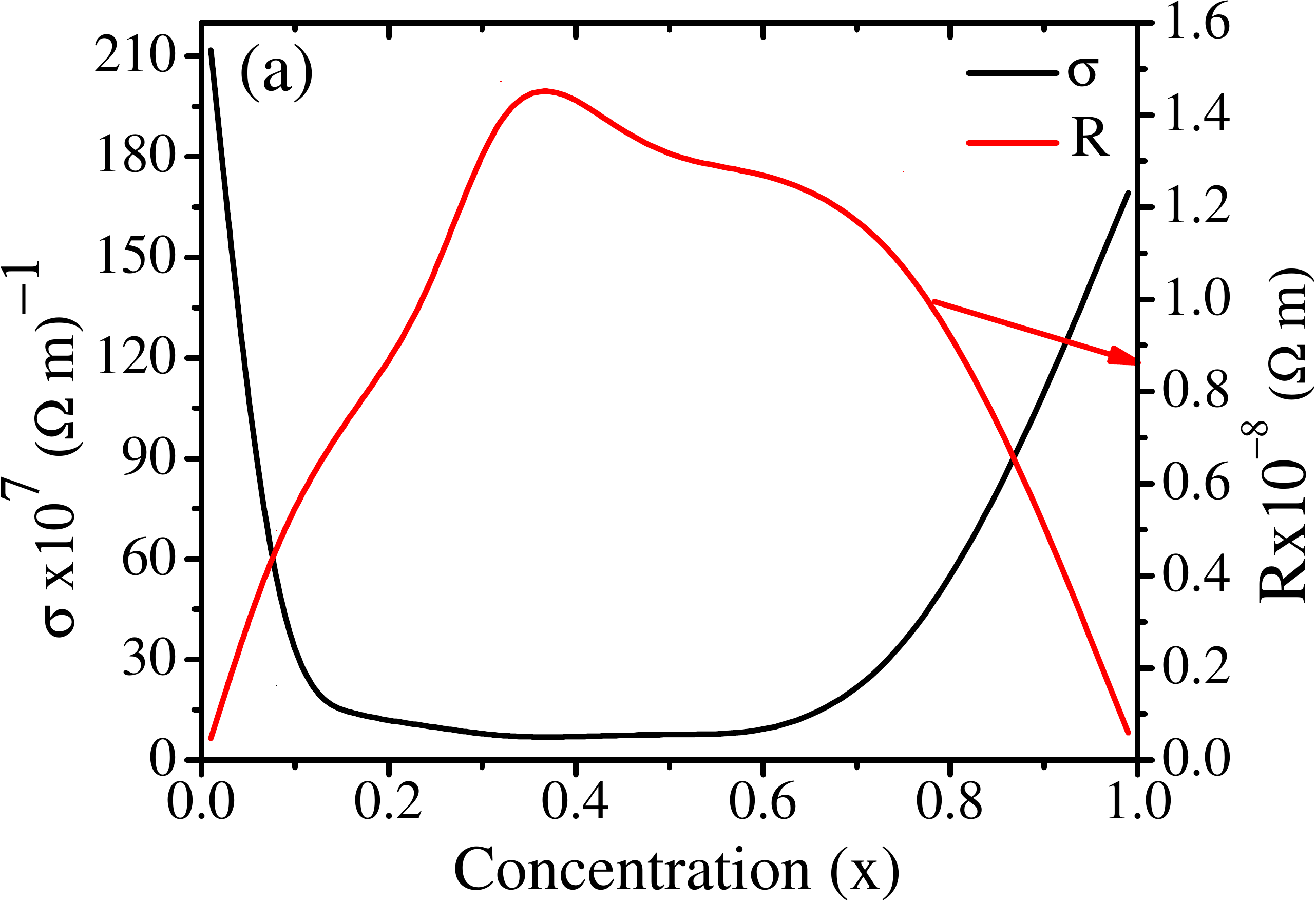}
     \includegraphics[width=3.9cm,height=3.4cm]{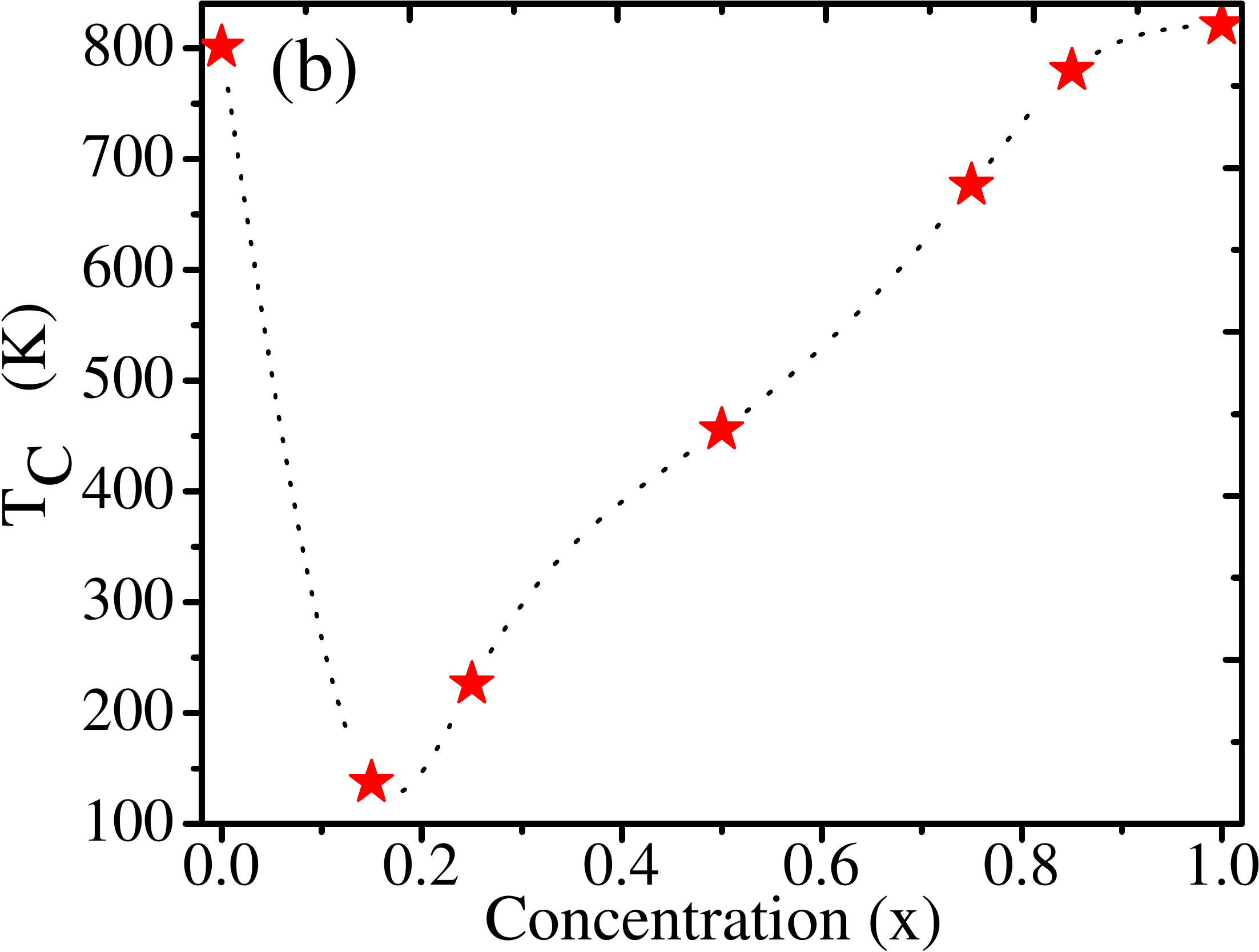}
    \caption{(a) Electrical conductivity ($\sigma$), Resistivity $R$ (scaling on the right hand side) and (b) Curie Temperature (T$_C$) of Cu$_2$MnAl$_{1-x}$Ga$_{x}$}
    \label{fig:cond-res}
\end{figure}

We show in Fig.\ref{fig:cond-res}(a) the electrical conductivity $\sigma$ and resistivity $R$ of Cu$_{2}$MnAl$_{1-x}$Ga$_{x}$ as obtained with the multiscattering Green function method. As the doping concentration of Ga increases, $\sigma$ decreases rather exponentially up to $x\sim 0.20$ and remained rather flat between $0.20 \leq x \leq 0.60$ before gradually increasing till $x=1.0$. In the resistivity, we observed opposite effect, but at the intermediate doping concentration, there is an apparent maximum occurring at $x\sim 0.4$, which is masked in the conductivity data. Hence, one can conclude that Cu$_{2}$MnAl$_{1-x}$Ga$_{x}$ has a maximum disorder at this optimal doping concentration. We further show in Fig.\ref{fig:cond-res}(b) the variation of $T_{C}$ with Ga doping concentration calculated within our mean-field approximation. At the Ga-deficit limit (Cu$_2$MnAl), T$_C\sim 798$ K then decreased almost exponentially to T$_C\sim 49$ K at 0.2 Ga doping concentration before systematically increasing rather monotonically to T$_C\sim 810$ K at the Ga-rich phase. We attribute the initial decrease of the T$_C$ at low Ga-doping to be due to the $sp$-defect states inducing antiferromagnetic coupling between the Mn-sites, which weakens the overall ferromagnetic exchange interaction in the system. The ferromagnetic exchange interaction, however, dominated as the concentration of the Ga-atom increased as evident from the systematic increase of the T$_C$ from $x=0.2$ to $x=1.0$ [c.f. Fig.~\ref{fig:cond-res}(b)]. We note, however, that our calculated T$_C$ values are higher than the experimental value T$_C$$\sim$ 630 K\cite{Oxley}. We attribute this discrepancy to the lack of magnetic percolation effect~\cite{Sato,PhysRevLett.93.137202,Fukushima} that is missing in our approximation. Even at this, magnitude-wise, the rather high T$_C$-value is a fingerprint that Cu$_{2}$MnAl$_{1-x}$Ga$_{x}$ could be a suitable candidate for making technological devices for high-temperature applications.

\begin{table}[htb]
\caption{The total ($m$), spin (m$_{s}$), and orbital (m$_{l}$) moment in units of $\mu_B$ for Cu and Mn in Cu$_{2}$MnAl$_{1-x}$Ga$_{x}$ at various doping concentration $x$ of Ga obtained within the first-principles, relativistic multiscattering KKR approach.} 
\label{tab:table1}
\begin{tabular}{cccccccc}
\hline \hline
 & \multicolumn{3}{c}{Cu} & \multicolumn{1}{c}{} & \multicolumn{3}{c}{Mn} \\ 
\cline{2-4}\cline{6-8}
$x$ & $m_s$ & $m_l$ & $m$ && $m_s$ & $m_l$ & $m$ \\ 
\hline
0.00  & 0.043 &  0.005 & 0.048   && 3.377 & 0.007 & 3.384\\
0.25  & 0.041 &  0.005 & 0.046   && 3.401 & 0.008 & 3.409 \\
0.50  & 0.041 &  0.000 & 0.041     && 3.451 & 0.000 & 3.451 \\
0.75  & 0.040 &  0.000 & 0.040   && 3.484 & 0.000 & 3.484 \\
0.85  & 0.043 &  0.004 & 0.075   && 3.505 & 0.008 & 3.513 \\
1.00  & 0.043 &  0.000 & 0.043   && 3.532 & 0.000 & 3.532  \\
\hline\hline
\end{tabular}
\end{table}

\subsection {X-ray Absorption Spectra and X-ray Magnetic Circular Dichroism}
\begin{figure}[t!]
    \centering
    \includegraphics[trim = 0mm 0mm 0mm 0mm,scale=0.20,keepaspectratio,clip=true]{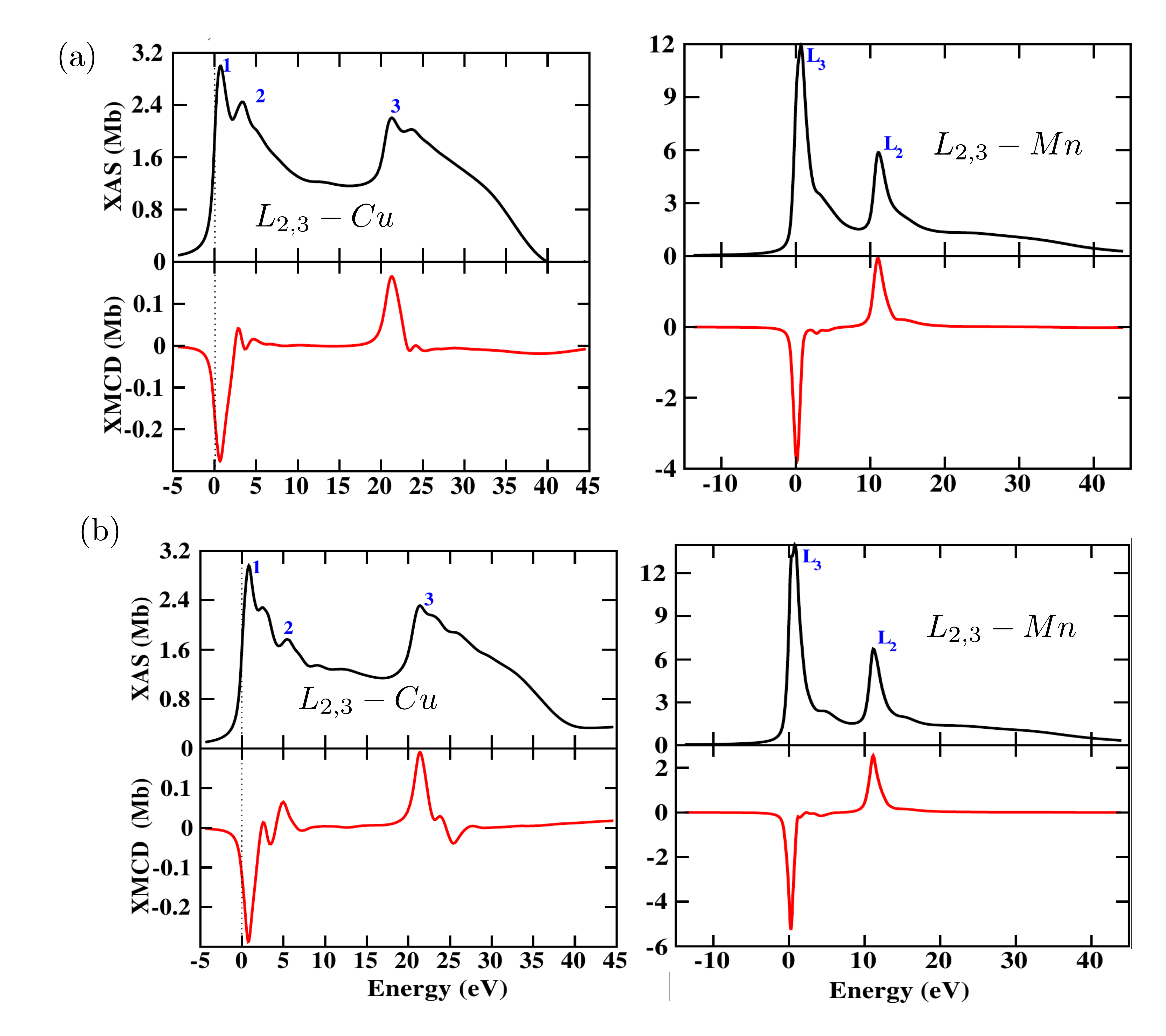}
    \caption{The X-ray absorption spectroscopy (XAS) (black line) and X-ray magnetic circular dichroism (XCMD) spectroscopy (red line) of Cu and Mn in (a) Cu$_{2}$MnAl and (b) Cu$_{2}$MnGa.}
    \label{fig:xas-xmcd}
\end{figure}

There are several theoretical methods to compute the X-ray absorption spectroscopy and X-ray magnetic circular dichroism spectra depending on the electronic signature of the material. For instance, the electron multiple scattering method\cite{PhysRevLett.70.694,Laan}, the configuration interaction approach, which is mostly used for metallic systems~\cite{Jo}, and the sum rules and multipole-moment techniques that are most suitable for heavier 3$d$ elements~\cite{Piamonteze,hedman2007x}. In our calculations, we adopted the sum rule approach within the framework of spin density functional theory. In  Fig.~\ref{fig:xas-xmcd}, we show the XAS and XMCD spectra of Cu and Mn at L$_{2,3}$ edges of Cu$_{2}$MnAl$_{1-x}$Ga$_{x}$ ($x=$0.0 and 1.0) obtained as described above at zero applied field.

Observe from Fig.~\ref{fig:xas-xmcd} that the X-ray absorption spectroscopy of Mn atoms is sharper with higher intensity as compared to that of the Cu atoms. In particular, the ratio of the intensities of the X-ray absorption between L$_{3}$ and L$_{2}$ edges in Mn is 2.4:1.0 while the ratio of the L$_{3}$ and L$_{2}$ edges of the X-ray absorption in Cu is 1.78:10, leading to a deviation of the branching ratio of 2:1 obtained from single-particle theory~\cite{hedman2007x,Schwitalla}. This discrepancy could be due to the lack of electron core-hole Coulomb and exchange correlations. As reported earlier, the $3d$-Mn with nearly empty states gives strong L$_{2,3}$ absorption peaks due to strong photon-electron and core-hole interactions~\cite{Krume}. In the X-ray absorption spectroscopy spectra of Cu, we observed three prominent peaks denoted by 1, 2, and 3 at 1.0, 4.5, and 21.0 eV, respectively. We attribute peaks 1 and 2 in the XAS to the L$_3$ and peak 3 to the L$_2$ edges. An ultraviolet photoemission spectroscopy measurement reported a maximum peak at $\sim$3.2 eV in the absorption edge, dominated mainly by Cu-3$d$ states~\cite{Brown}. The X-ray magnetic circular dichroism spectrum of Cu at the L$_{2,3}$ edges consists of a negative peak at 1.0 eV, two positive peaks at 4.5 eV and 21.0 eV, followed by a small negative peak at 25.0 eV (c.f. Fig.\ref{fig:xas-xmcd}). 

Comparing the peaks of the X-ray absorption spectroscopy with the corresponding $d$-character of the density of states of Cu/Mn, we confirm a strong hybridization with the nearest Mn atoms. Most of the $d$-electrons of Cu atoms are localized as such, their contribution to the XAS is negligible in comparison to that of the Mn-$d$ orbitals. Hence, the Mn-$d$ states play a vital role in the formation of the XMCD spectrum due to their relatively large spin and orbital moments. We note that our results are in qualitative agreement with the experimentally measured XMCD spectrum at Cu/Mn L$_{2,3}$ edges~\cite{Krume,Telling}. The first sharp peak denoted by 1 is due to transitions to the minority ($\downarrow$) spin $t_{2g}/e_{g}$ states of Cu/Mn atoms that form a small peak in the density of states at $\sim$1.0 eV above the Fermi level [c.f. Fig.~\ref{fig:CMA-dos}(a)]. As shown in Fig.\ref{fig:xas-xmcd}(b) the next shoulder after peak 1 at 2.0 eV is due to the hybridization between Cu($\downarrow$)-$s$, Cu($\downarrow$)-$d$ and Al($\downarrow$)-$p$ states. A peak 2 at 3.5 eV is a result of an overlap between Cu($\downarrow$) and Al($\downarrow$) states.

\section{Conclusion}
The electronic structure and X-ray spectra of Cu$_2$MnAl$_{1–x}$Ga$_x$ have been studied using a first-principles, relativistic multiscattering Green function technique. We have analyzed the variation of the partial magnetic moment with an $sp$-electron doping at Al sites. Our calculations show that as the concentration of Ga doping increases, the exchange splitting on the Mn-site is enhanced, a signature of possible antiferromagnetic direct exchange. Our resistivity data show that the dopant has the maximum effect at the doping concentration x  0.4. In relation to the exchange coupling, we estimate the Curie temperature within a mean field approximation. The TC initially decreased to 49 $K$ at 0.2 Ga-doping concentration before systematically increasing to a maximum value of 810 $K$ at the Ga-rich phase. We attributed this to the weakening of the ferromagnetic exchange interactions due to the induced antiferromagnetic coupling by the $sp$-defect states between the Mn-sites. Overall, the calculated T$_C$ is overestimated as compared to experimental value due to the lack of magnetic percolation effect. However, the rather high T$_C$-value is a signature that Cu$_2$MnAl$_{1–x}$Ga$_x$ could be a suitable candidate for making technological devices for high-temperature applications. The XAS and XMCD spectra are calculated at the Cu and Mn sites for both Cu$_2$MnAl and Cu$_2$MnGa phases, which are in qualitative agreement with the experimental ones. 

\section{Acknowledgment}
C.E.E. acknowledges the support of the University of Port Harcourt and the Government of Ebonyi, Nigeria. D.P.R. acknowledges DST-RFBR, Project Ref. No. INT/RUS/RFBR/P-264.
\ifx\mcitethebibliography\mciteundefinedmacro
\PackageError{unsrtM.bst}{mciteplus.sty has not been loaded}
{This bibstyle requires the use of the mciteplus package.}\fi

\end{document}